\documentclass[aps,preprint,letterpaper,superscriptaddress,longbibliography]{revtex4-2}
\usepackage[justification=raggedright,singlelinecheck=false,labelfont=bf,labelsep=period]{caption,subcaption}
\usepackage{braket}
\usepackage{siunitx}
\usepackage{comment}
\usepackage{nameref}
\usepackage[version=4]{mhchem}

\usepackage{scalerel}
\usepackage{tikz}
\usetikzlibrary{svg.path}
\definecolor{orcidlogocol}{HTML}{A6CE39}
\tikzset{
  orcidlogo/.pic={
    \fill[orcidlogocol] svg{M256,128c0,70.7-57.3,128-128,128C57.3,256,0,198.7,0,128C0,57.3,57.3,0,128,0C198.7,0,256,57.3,256,128z};
    \fill[white] svg{M86.3,186.2H70.9V79.1h15.4v48.4V186.2z}
                 svg{M108.9,79.1h41.6c39.6,0,57,28.3,57,53.6c0,27.5-21.5,53.6-56.8,53.6h-41.8V79.1z M124.3,172.4h24.5c34.9,0,42.9-26.5,42.9-39.7c0-21.5-13.7-39.7-43.7-39.7h-23.7V172.4z}
                 svg{M88.7,56.8c0,5.5-4.5,10.1-10.1,10.1c-5.6,0-10.1-4.6-10.1-10.1c0-5.6,4.5-10.1,10.1-10.1C84.2,46.7,88.7,51.3,88.7,56.8z};}}
\newcommand\orcidicon[1]{\href{https://orcid.org/#1}{\mbox{\scalerel*{
\begin{tikzpicture}[yscale=-1,transform shape]
\pic{orcidlogo};
\end{tikzpicture}
}{|}}}}

\usepackage{etoolbox}

\makeatletter
\pretocmd\frontmatter@thefootnote{\color{blue}}{}{}

\usepackage{dcolumn} 
\usepackage{titlesec}
\titlespacing*{\section}
{0pt}{4ex}{1.3ex}
\titlespacing*{\section}
{0pt}{4ex}{0ex}

\titleformat{\section}
  {\bfseries\MakeUppercase}{\thesection\raggedright}{1em}{}
  \titleformat{\section}
 {\bfseries}{\thesection\raggedright}{1em}{}

\usepackage{xstring}
\usepackage{xspace}

\usepackage[utf8]{inputenc}
\usepackage[english]{babel}
\usepackage{gensymb}
\usepackage{mathtools}

\usepackage{float}
\usepackage{graphicx}

\usepackage{url}
\usepackage[utf8]{inputenc}
\usepackage[T1]{fontenc}
\usepackage{textcomp}
\usepackage{gensymb}

\newcommand{\angstrom}{\mbox{\normalfont\AA}}

\usepackage[colorlinks,citecolor=red,urlcolor=blue,bookmarks=false,hypertexnames=true]{hyperref}
\usepackage{cleveref}

\begin{document}

\title{Directional atomic layer etching of MgO-doped lithium niobate using Br-based plasma}


\author{Ivy I. Chen~\orcidicon{0009-0009-0942-5658}}

\affiliation{%
Division of Engineering and Applied Science, California Institute of Technology, Pasadena, California 91125, USA
}%

\author{Mariya Ezzy~\orcidicon{0009-0002-4791-0995}}

\affiliation{%
Division of Engineering and Applied Science, California Institute of Technology, Pasadena, California 91125, USA
}%

\author{Emily Hsue-Chi Shi~\orcidicon{0009-0003-2807-7106}}

\affiliation{%
Division of Chemistry and Chemical Engineering, California Institute of Technology, Pasadena, California 91125, USA
}%

\author{Clifford F. Frez}

\affiliation{%
Jet Propulsion Laboratory, California Institute of Technology, Pasadena, California 91109, USA
}%

\author{Suraj~\orcidicon{0000-0002-5724-3746}}

\affiliation{%
Jet Propulsion Laboratory, California Institute of Technology, Pasadena, California 91109, USA
}%

\author{Lin Yi}

\affiliation{%
Jet Propulsion Laboratory, California Institute of Technology, Pasadena, California 91109, USA
}%

\author{Mahmood Bagheri}

\affiliation{%
Jet Propulsion Laboratory, California Institute of Technology, Pasadena, California 91109, USA
}%

\author{James R. Renzas}
\affiliation{%
College of Engineering, University of Nevada, Reno, Reno, Nevada 89557, USA
}%

\author{Alireza Marandi}
\affiliation{%
Department of Electrical Engineering, California Institute of Technology, Pasadena, California 91125, USA
}%

\author{Frank Greer}
\affiliation{%
Jet Propulsion Laboratory, California Institute of Technology, Pasadena, California 91109, USA
}%

\author{Austin J. Minnich~\orcidicon{0000-0002-9671-9540} }
 \email{aminnich@caltech.edu}
\affiliation{%
Division of Engineering and Applied Science, California Institute of Technology, Pasadena, California 91125, USA
}%

\date{\today}
\begin{abstract}
Lithium niobate (LiNbO$_3$, LN) is a nonlinear optical material of high interest for integrated photonics with applications ranging from optical communications to quantum information processing. The performance of on-chip devices based on thin-film lithium niobate (TFLN) is presently limited by fabrication imperfections such as sidewall surface roughness and geometry inhomogeneities over the chip. Atomic layer etching (ALE) could potentially be used to overcome these difficulties. Although an isotropic ALE process for LN has been reported, performing LN fabrication completely with ALE faces several challenges, including the lack of a directional ALE process for pattern transfer and the redeposition of involatile compounds.  Here, we report a directional ALE process for LN consisting of sequential exposures of HBr/BCl$_3$/Ar plasma for surface modification and Ar plasma for removal. The HBr chemistry is found to decrease redeposition compared to F- and Cl-based plasmas, which we attribute to the higher vapor pressures of Br-based products. A grating pattern etched entirely by the process (total etch depth of 220 nm) exhibits no aspect ratio dependent etching (ARDE) down to the smallest tested gap of 150 nm, in contrast to ion milling in which ARDE manifests even at 300 nm gaps for the same etch depth. The HBr plasma chemistry is also found to support an isotropic process consisting of sequential exposures of H$_2$ plasma and HBr/BCl$_3$/Ar plasma. These processes could be used together to perform the complete fabrication process for TFLN devices, eliminating imperfections arising from ion milling.
\end{abstract}
\maketitle
\newpage

\section{Introduction}

Lithium niobate (LiNbO$_3$ or LN) is an optical material of long-standing interest for nonlinear optics owing to its wide transparency window (350-5200 nm), high electro-optic coeffiecient ($r_{33}=30.9$ pm/V), high second-order nonlinear susceptibility ($d_{33}=-27$ pm/V at 1060 nm) and ferroelectric properties. \cite{Weis1985_LN_properties, wong2002_LN_properties, kong2020LN, volk2008LN, arizmendi2004LNphotonic} In the past decade, nanophotonics in TFLN has experienced rapid advancement due to developments in thin-film substrate production and optimized etching processes for low-loss waveguides. \cite{Zhang:17, Zhu:21} These advances enable a variety of integrated photonics applications including optical computing and quantum information technologies. \cite{Guo2022,Hu2025, Marandi_2022_squeezed_states, Finco2024} 

Presently, fabrication imperfections limit the performance and scalability of TFLN as an integrated photonics platform. For instance, the efficiency of second-order effects such as sum- and difference- frequency generation is sensitive to nanometer-scale device geometry variations in top width and height of waveguides \cite{Zhao2023_QPM, Kuo:22, fejer2002quasi}; a simulation study found that the frequency doubling efficiency in a 5-mm-long device is reduced by 50\% for a 2.2 nm variation in waveguide thickness. \cite{Kuo:22} A state-of-the-art TFLN chip is typically patterned with hydrogen silsesquioxane (HSQ) and dry etched using physical Ar$^+$ milling, which exhibits limitations such as low etch selectivity with HSQ, non-vertical sidewalls, sidewall surface roughness, redeposition of LN, and lack of nanometer-level etch depth control. \cite{Zhu:21, Ar_LN_Etch_Clean} Some of these limitations have been remedied through targeted approaches; for instance, a TFLN chip typically contains devices with systematically varied design parameters to increase the likelihood that at least one device meets the design specifications. \cite{Hwang:23} Other methods to address fabrication limitations include thermally tuning chips. \cite{Xin2025_TFLN} Nevertheless, there is a need for improved nanofabrication methods to accelerate the development and integration of TFLN devices into larger-scale photonic integrated circuits. 

To this end, atomic layer etching (ALE) has emerged as a promising nanofabrication technique for fabricating TFLN devices. Compared to conventional etching techniques, ALE imparts less surface damage and enables intrinsic wafer-scale uniformity in etch depth.  \cite{10.1116/6.0003899, coatings11030268, MIN201382, 10.1116/1.4816321, mahuli2025improvinglifetimealuminumbasedsuperconducting, doi:10.1021/acs.chemmater.4c01606, 10.1116/6.0003263, doi:10.1021/acssuschemeng.2c05186, 10.1116/6.0003593} ALE typically consists of two sequential, self-limiting surface reactions which together yield sub-angstrom to single-nanometer-scale etching per cycle.  \cite{kanarik_rethinking_etch, kanarik_ALE_overview} We recently demonstrated isotropic ALE processes for LN using fluorine and chlorine chemistries. \cite{Chen2024} However, both processes exhibited surface roughening of polished samples at the process temperature of \SI{0}{\degreeCelsius}, which was attributed to redeposition of lithium and magnesium halides.  Traditionally, redeposition has been addressed using heavier halogens \cite{Flanders_InEtchWithI} and elevated temperatures \cite{tsou1993ITOHBr} to increase the volatility of the etch products. The vapor pressures of lithium and magnesium halides are generally higher for heavier halogens like Br compared to lighter ones like F and Cl. \cite{doi:10.1021/ie50448a022, Shibata_2014} However, to the best of our knowledge, neither continuous nor self-limiting etching processes based on HBr chemistry have been investigated for LN.

Here, we report both a directional ALE process for LN using an HBr-based plasma chemistry. The process consists of sequential exposures of an HBr-containing plasma followed by an Ar plasma for removal by low-energy ion sputtering. We obtain an etch per cycle (EPC) of $1.04 \pm 0.01$ nm/cycle with a synergy of $85\%$ at \SI{0}{\degreeCelsius}. At a process temperature of \SI{200}{\degreeCelsius}, the synergy decreases but the surface remains atomically smooth after 50 cycles of ALE, in contrast to a Cl-based plasma chemistry which produced rougher surfaces.  A TFLN grating structure etched completely by ALE (total etch depth of 220 nm) was fabricated and found to exhibit aspect ratio independent etching behavior down to the smallest tested gap of 150 nm. Additionally, an isotropic HBr-based ALE process at \SI{0}{\degreeCelsius} is reported. Together, these processes could be used for the complete fabrication flow of TFLN devices, overcoming the limitations of physical dry etching.

\section{Methods}

The samples in this study consisted of bulk 3-inch x-cut 5\% mol MgO-doped single crystal LN wafers (G \& H Photonics) which were diced into 7 mm $\times$ 7 mm substrates using a Disco DAD 321 dicing saw and cleaned by sonication in AZ NMP Rinse, acetone, and isopropyl alcohol.
Two process temperatures, $\SI{0}{\degreeCelsius}$ and \SI{200}{\degreeCelsius}, were tested to evaluate whether nonvolatile etch products potentially present at \SI{0}{\degreeCelsius} could be volatilized at \SI{200}{\degreeCelsius}. The process temperature was measured by the table thermometer. For the \SI{0}{\degreeCelsius} processes, the table was cooled using liquid nitrogen. The samples were patterned with AZ5214E IR photoresist following our previous report. \cite{Chen2024} For \SI{200}{\degreeCelsius} etches, the samples were patterned with hydrogen silsesquioxane (HSQ) using electron beam lithography (Raith EBPG 5200) to ensure mask compatibility. The samples were etched in an Oxford Instruments PlasmaPro 100 Cobra system configured for ALE. 

\begin{figure}
    \centering
    {\includegraphics[width=\textwidth]{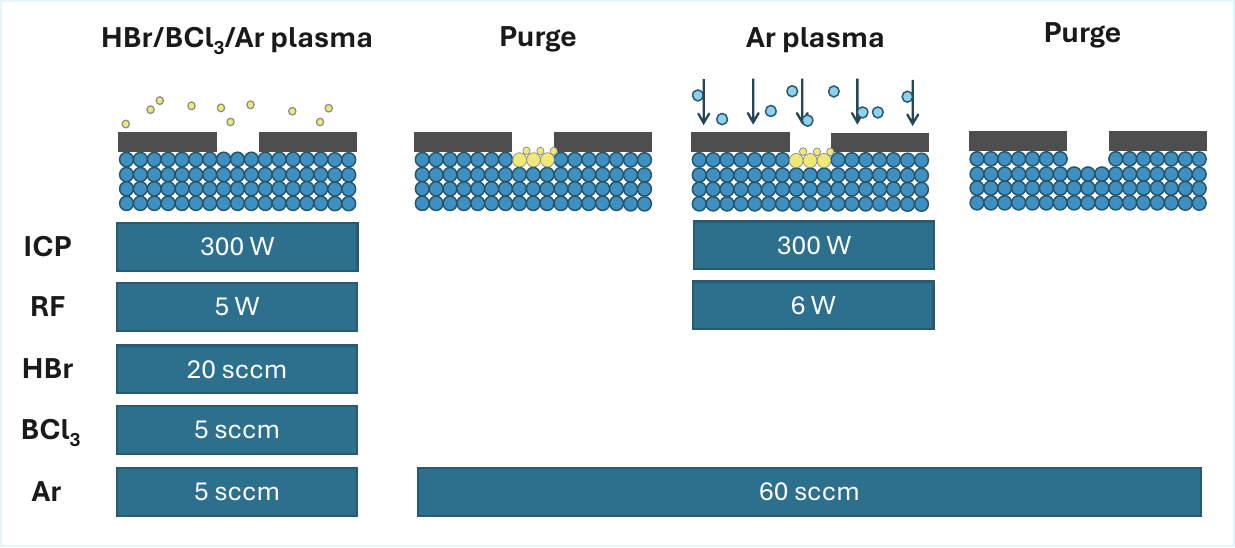}
    }
    \caption{Schematic of the directional ALE process for LN. First, an HBr/BCl$_3$/Ar plasma exposure modifies the surface, followed by a purge. A low-power Ar plasma follows to sputter off the modified surface. A final purge completes the ALE cycle.}
    \label{fig:ALE_recipe}
\end{figure}

A schematic of the directional process is shown in \Cref{fig:ALE_recipe}. The ALE recipe consists of a 40-second HBr/BCl$_3$/Ar plasma exposure (300 W ICP power, 5 W RIE power, 30 V DC bias, 20 sccm HBr, 5 sccm BCl$_3$, 5 sccm Ar) followed by a 60-second Ar plasma exposure (300 W ICP power, 6 W RIE power, 45 V DC bias, 60 sccm Ar). The HBr/BCl$_3$/Ar plasma step utilized a weak RIE power to allow for reliable striking of the plasma. The addition of BCl$_3$ to the HBr gas mixture was used to facilitate the etching of the oxygen-containing components of the compound.  The 5 second purge times with 60 sccm Ar were used between plasma half-steps. The chamber pressure was set at 10 mTorr. 

To measure saturation curves for the directional process, the chamber pressure, table temperature, and ICP power were kept constant at 10 mTorr, \SI{0}{\degreeCelsius}, and 300 W respectively, while the exposure time for each half-step was varied. HBr/BCl$_3$/Ar plasma exposure time was varied from 0 to 50 seconds with Ar plasma held at 60 seconds, and Ar plasma exposure time was varied from 0 to 80 seconds with HBr/BCl$_3$/Ar plasma exposure held at 40 seconds. Prior to introducing the sample into the chamber for etching, a blank Si wafer was introduced and the etching chamber was cleaned with a 15-minute Ar plasma with 1500 W ICP and 100 W RF power followed by a 15-minute O$_2$/SF$_6$ plasma with the same power parameters. After the sample was loaded into the chamber, a 3-minute wait time before processing allowed the sample to thermally equilibrate with the table. To investigate whether higher process temperatures would yield smoother surfaces and less redeposition, the same processes were carried out at a table temperature of \SI{200}{\degreeCelsius}. 


In addition to the directional process, an isotropic process based on HBr-containing plasmas was also identified. This process consisted of exposures of H$_2$ plasma followed by HBr/BCl$_3$/Ar plasma. The HBr-based plasma used in the directional process replaces the second step in the isotropic process reported in Ref. \cite{Chen2024}. The plasma settings of the H$_2$ (300 W ICP power, 20 W RIE power, 200 V DC bias, 60 sccm H$_2$) were slightly changed to achieve the same DC bias in Ref. \cite{Chen2024}. Synergy, etch rate, and surface roughness were measured for the isotropic process at a process temperature of \SI{0}{\degreeCelsius} only.

After etching, photoresist patterned samples were cleaned by sonication in AZ NMP Rinse, acetone, and isopropyl alcohol. For HSQ patterned samples, a buffered HF dip (Buffer HF improved, Transene Company, Inc.) at room temperature was used to remove the mask after etching. Etch per cycle (EPC) was calculated by measuring the etch depth for a processed sample and dividing it by the total number of cycles. AFM scans were performed on a Bruker Dimension Icon atomic force microscope (AFM) to measure total etch depth and surface roughness. The total etch depth was measured using $2.5 \times 10$ $\mu$m$^2$ AFM scan with the scan rate set to 0.5 Hz. The step profile was averaged over the entire scan using Nanoscope Analysis 1.9 software to obtain the etch depth. RMS surface roughness of a reference TFLN Ar$^+$ milled waveguide sample and power spectral density (PSD) scans were obtained over a 50 $\times$ 50 nm$^2$ area with a 0.5 Hz scan rate. Waveguide sidewall slope on measured TFLN samples and sample tilt from all AFM scans were removed via quadratic plane fit.

The synergy, $S$, as defined by Ref.~\cite{kanarik_ALE_synergy}, quantitatively compares the etch depth using only individual steps of the ALE cycle to the etching obtained with the full etch cycle. It is defined as $S=(1 - (\alpha+\beta)/EPC)\times 100$, where $\alpha$ and $\beta$ are the etch rate of the half-cycles, and $EPC$ is the etch rate of the full cycle. The synergy of the directional and isotropic process was measured with atomic force microscopy (AFM) after 20 cycles of ALE.

 
X-ray photoelectron spectroscopy (XPS) analysis was performed using a Kratos Axis Ultra x-ray photoelectron spectrometer using a monochromatic Al K$\alpha$ source. A neutralizer was used to compensate for sample charging during analysis. The resulting data was analyzed in CASA-XPS from Casa Software Ltd. For each sample, we collected the carbon C1s, oxygen O1s, niobium Nb3d$_{5/2}$ and Nb3d$_{3/2}$, bromine Br3d$_{5/2}$ and Br3d$_{3/2}$, niobium Nb4s, lithium Li1s, and magnesium Mg2p peaks. The Nb3d$_{5/2}$ peaks were used to calibrate peak positions using the NIST reference for LN (207.3 eV) \cite{database2000nist}. We fit the data using a Shirley-type background subtraction with offsets and peak fitting routines from Refs.~\cite{2002_LN_H2_XPS, gruenke2023LNXPS, LN_XPS_Nb}. 

\section{Results}
\subsection{ALE process characteristics}

\begin{figure}
    \centering
    
{\includegraphics[width = 0.75\textwidth]{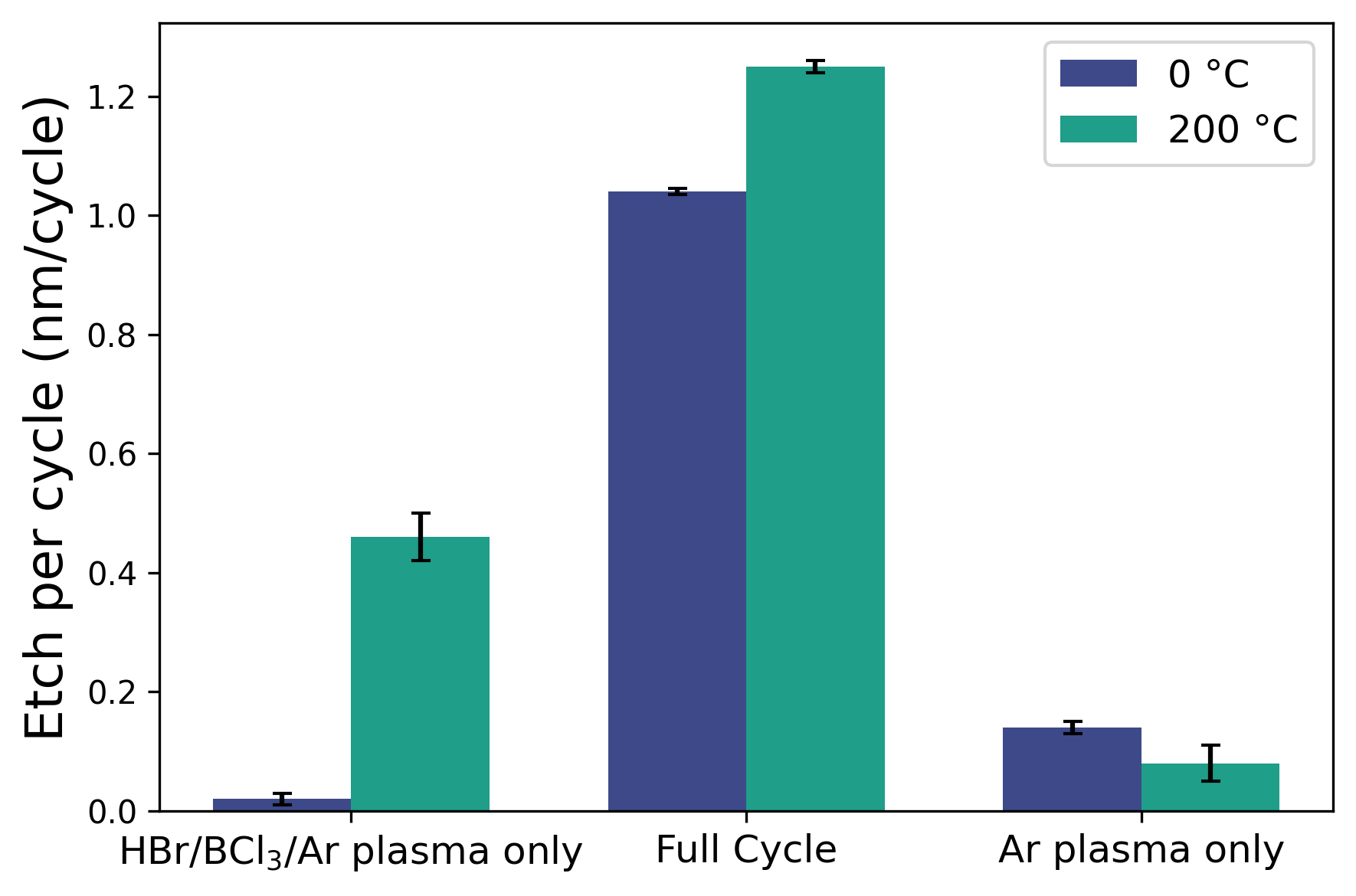}
}
\caption{ Etch per cycle for the half cycles and full cycle of the directional ALE process at \SI{0}{\degreeCelsius} and \SI{200}{\degreeCelsius}.
 The synergy at \SI{0}{\degreeCelsius} (\SI{200}{\degreeCelsius}) is 84.6\% (56.8\%).}
\label{fig:synergy} 
\end{figure}


We first discuss the synergy characteristics of the directional ALE recipe at the two temperatures. \Cref{fig:synergy} shows the EPCs of the individual half-steps and the full cycle of the process. The half-cycle EPCs were measured as $0.02\pm0.01$ nm/cycle and $0.14\pm 0.01$ nm/cycle for the HBr/BCl$_3$/Ar plasma half step and Ar half step, respectively. When using both steps sequentially, an etch rate of 1.04$\pm$0.02 nm/cycle is observed. These data yield a synergy of $S=84.6$\%. 

At \SI{200}{\degreeCelsius}, the etch rate and synergy were measured to be $1.25 \pm 0.01$ nm/cycle and 56.8\%, respectively. The synergy at this temperature is lower compared to \SI{0}{\degreeCelsius} due to an increase in EPC for the  HBr/BCl$_3$/Ar plasma half step, with the EPC increasing to  $0.46\pm0.04$ nm/cycle from 0.02 nm/cycle.  The increase in EPC for the HBr/BCl$_3$/Ar half-step is attributed to increased spontaneous etching at higher temperatures. A possible spontaneous reaction mechanism at \SI{200}{\degreeCelsius} could be a conversion of the metal oxide to  B$_2$O$_3$ with BCl$_3$, which then gets spontaneously etched with BCl$_3$ via the reaction B$_2$O$_3$ + BCl$_3$(g) → B$_3$O$_3$Cl$_3$(g), as in thermal ALE of various metal oxides.  \cite{Cano2022ThermalALEalumina, Gertsch2023_thermalALEVo2, Johnson_thermalALE_WO3} At room temperature, the condition for many industrial etch processes, the etch characteristics would likely be most similar to the \SI{0}{\degreeCelsius} results.

\begin{figure}
    \centering
    
{\includegraphics[width = \textwidth]{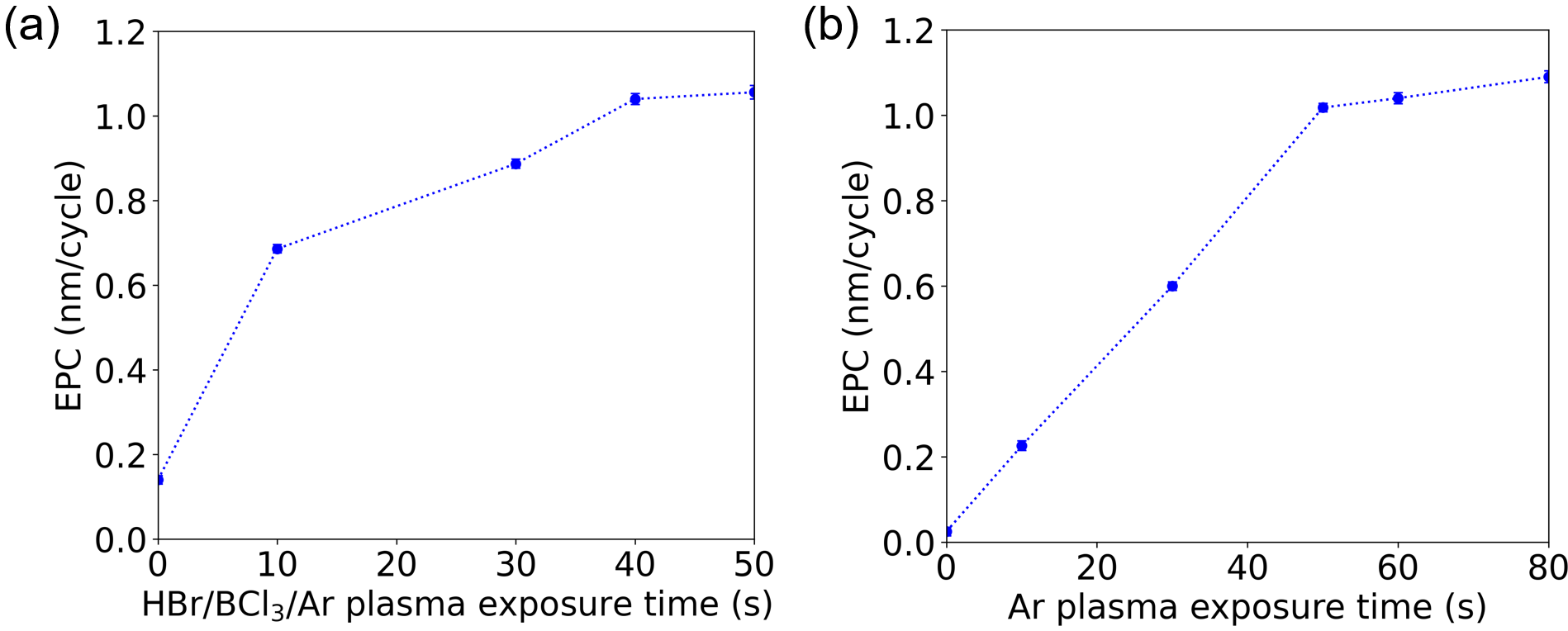}
    \phantomsubcaption\label{fig:HBr_selflimit}
    \phantomsubcaption\label{fig:Ar_selflimit}
}
\caption{(a) EPC versus HBr/BCl$_3$/Ar plasma exposure time with Ar plasma exposure time fixed at 60 s. (b) EPC versus Ar plasma exposure time with HBr/BCl$_3$/Ar plasma exposure time fixed at 40 s.  All experiments were performed at \SI{0}{\degreeCelsius}. Dashed lines are guides to the eye.
}
\label{fig:self-limiting} 
\end{figure}

Next, we characterized the self-limiting nature of the process by measuring the saturation curves for each half-cycle. Due to experimental limitations, this measurement was performed at  \SI{0}{\degreeCelsius} only. \Cref{fig:HBr_selflimit} shows the EPC for which the Ar plasma half step is held constant at 60 seconds while the HBr/BCl$_3$/Ar plasma exposure time is varied from 0 to 50 seconds. Saturation is observed to occur at $1.04 \pm 0.01$ nm/cycle after 40 seconds of  HBr/BCl$_3$/Ar plasma exposure time. The complementary case in which the HBr/BCl$_3$/Ar plasma exposure time is held constant at 40 seconds while the Ar plasma exposure time is varied from 0 to 80 seconds is shown in \Cref{fig:Ar_selflimit}. The EPC plateaus with increasing Ar plasma exposure time with a soft saturation of 0.23 \angstrom/cycle per 10 seconds of additional Ar plasma exposure. Soft saturation is hypothesized to occur due to the presence of a concentration gradient of bromine into the LN film after HBr/BCl$_3$/Ar plasma exposure. By increasing the Ar plasma exposure time, more of the chemically modified surface is removed, resulting in a soft-saturating curve.

We also identified an isotropic ALE process at \SI{0}{\degreeCelsius} with  HBr/BCl$_3$/Ar plasma using the same process sequence as our previous report \cite{Chen2024}, with the halogen plasma modification step replaced with the HBr/BCl$_3$/Ar plasma. This process yielded an etch rate of $1.43 \pm 0.02$ nm/cycle with a synergy of 97.9\%. The resulting RMS surface roughness after 20 cycles of etching is $R_q = 0.77 \pm 0.07$ nm, smoother than the reported roughness values of the previously reported Cl$_2$/BCl$_3$ chemistry of $R_q = 0.90 \pm 0.46$ nm with an EPC of $1.65 \pm 0.03$ nm/cycle \cite{Chen2024}. However, the HBr-etched surface is rougher than that obtained with the previously reported isotropic recipe using SF$_6$/Ar chemistry, which exhibited a resulting RMS surface roughness of $ 0.57 \pm 0.18$ nm after 20 cycles with a higher EPC of $1.59 \pm 0.02$ nm/cycle \cite{Chen2024}. However, it is noted that no optimization was performed to improve the surface roughness of the HBr-based isotropic process. 

The higher EPC in the isotropic etch compared to the directional value is attributed to the high DC bias (200 V) in the H$_2$ plasma modification step in the isotropic case, which creates a thicker modified layer per cycle and thus increases the EPC. All subsequent results are for the directional process.


\begin{figure}
    \centering
    
{\includegraphics[width = 450pt]{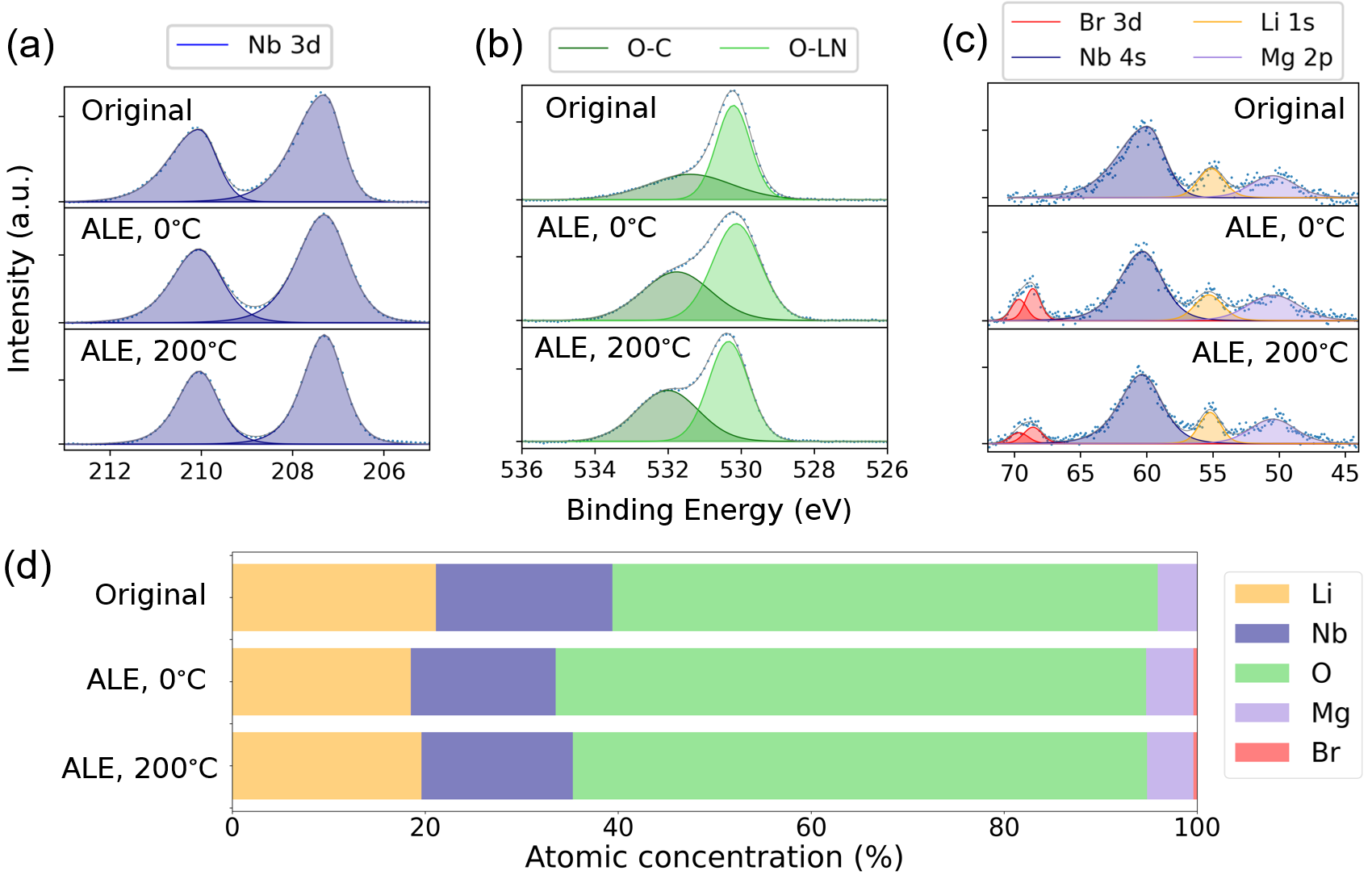}
    \phantomsubcaption\label{fig:directionalXPSa}
    \phantomsubcaption\label{fig:directionalXPSb}
    \phantomsubcaption\label{fig:directionalXPSc}
    \phantomsubcaption\label{fig:directionalXPS_atomic}
    
}
\caption{Surface XPS spectra of (a) Nb3d, (b) O1s, and (c) Br3d, Nb4s, Li1s, and Mg2p. The spectra are (top) original (middle) directional-ALE etched LN over 50 cycles at \SI{0}{\degreeCelsius} and (bottom) directional-ALE etched LN over 50 cycles at \SI{200}{\degreeCelsius}. (d) Atomic concentrations for original, 50 cycles directional ALE etched at \SI{0}{\degreeCelsius}, and 50 cycles directional ALE etched at \SI{200}{\degreeCelsius}. The atomic percentages are listed in \Cref{tab:xps_table}.
}
\label{fig:directionalxps} 
\end{figure}


\subsection{Surface composition and morphology}

We next characterize the surface chemical composition of bulk LN before and after 50 cycles of ALE  using XPS. The Nb3d$_{5/2}$ peak at 207.3 eV is used as a reference. \cite{database2000nist} In \Cref{fig:directionalXPSa,fig:directionalXPSb,fig:directionalXPSc}, we show the core levels of Nb3d, O1s; and Nb4s, Li1s, and Mg2p, respectively. No chlorine peaks were detected in the  Cl2p region of 195-200 eV.  For the control sample without any processing, we find the peak splitting of Nb3d to be 2.74 eV, in agreement with other studies. \cite{database2000nist} In \Cref{fig:directionalXPSb}, we report the O1s spectra with two subpeaks at  530.2 and 531.4 eV, corresponding to metal oxide and O-C bonds, respectively. \cite{chastain1992XPShandbook} In \Cref{fig:directionalXPSc}, we report the Nb4s, Li1s, and Mg2p spectra at 60.0 eV, 55.0 eV, and 50.1 eV, respectively (values are for original bulk LN). After ALE, we observe additional peaks corresponding to the Br 3d$_{3/2}$ and 3d$_{5/2}$ core levels at 68.6 and 69.6 eV, respectively.


\Cref{fig:directionalXPS_atomic} displays the histogram of the atomic concentrations obtained at different process temperatures. In \Cref{tab:xps_table}, we report the atomic concentrations of Nb, Li, Mg, O, and Br. The percentages were normalized by the surface adventitious carbon content (around 10-20\% depending on the sample), and the uncertainties in all atomic concentrations were obtained by propagating fit uncertainties derived from Monte Carlo simulations performed in CasaXPS. We note that there is a systematic difference between the calculated Li/Nb ratio and Mg concentration and the expected values of 0.95-0.97 and 5\% \cite{volk2008LN,Abrahams1986LN}, respectively. We attribute these differences due to the smaller relative sensitivity factors (RSF) of the Li1s and Mg2p orbitals (0.025 and 0.168 respectively), giving larger uncertainties for the atomic concentration compared to other elements. However, the presence of non-volatile etch products can be identified by comparing the Br (RSF = 1.06) surface concentration of the various samples. After 50 cycles at \SI{0}{\degreeCelsius}, the surface bromine concentration is found to be $0.50\pm 0.04 \%$. This concentration decreases to $0.35 \pm 0.06 \%$ for 50 cycles at \SI{200}{\degreeCelsius}, indicating that the higher process temperature has resulted in a decrease of Br etch products remaining on the surface. Further insight into the chemical mechanisms could be obtained using angle-resolved XPS to gain surface sensitivity and thermal desorption spectroscopy to identify the compounds present on the surface.


\begin{table}
\caption{\label{tab:xps_table} Atomic percentages for the fitted XPS data.}
\begin{ruledtabular}
\begin{tabular}{cccccc}
 Sample & Nb (\%) & O (\%) & Li (\%) & Mg(\%) & Br (\%)\\
\hline
Untreated   &  $18.34 \pm 0.78$    &    $ 56.50 \pm 2.43 $      &        $ 21.11 \pm 1.76 $    &  $4.06 \pm 0.77 $  & $0$\\
50 cycles ALE at \SI{0}{\degreeCelsius}  &  $14.95 \pm 0.27$ &   $61.16 \pm 1.10$  &      $18.48 \pm 1.08$   &  $4.90 \pm 0.61$  &  $0.50 \pm 0.04$\\
50 cycles ALE at \SI{200}{\degreeCelsius}  &  $15.83 \pm 0.28$ &  $59.47 \pm 1.05$   &     $19.60 \pm 1.00$     &  $4.76 \pm 0.58$ &  $0.35 \pm 0.06$
\end{tabular}
\end{ruledtabular}
\end{table}

\begin{figure}
    \centering
    
{\includegraphics[width=\textwidth]{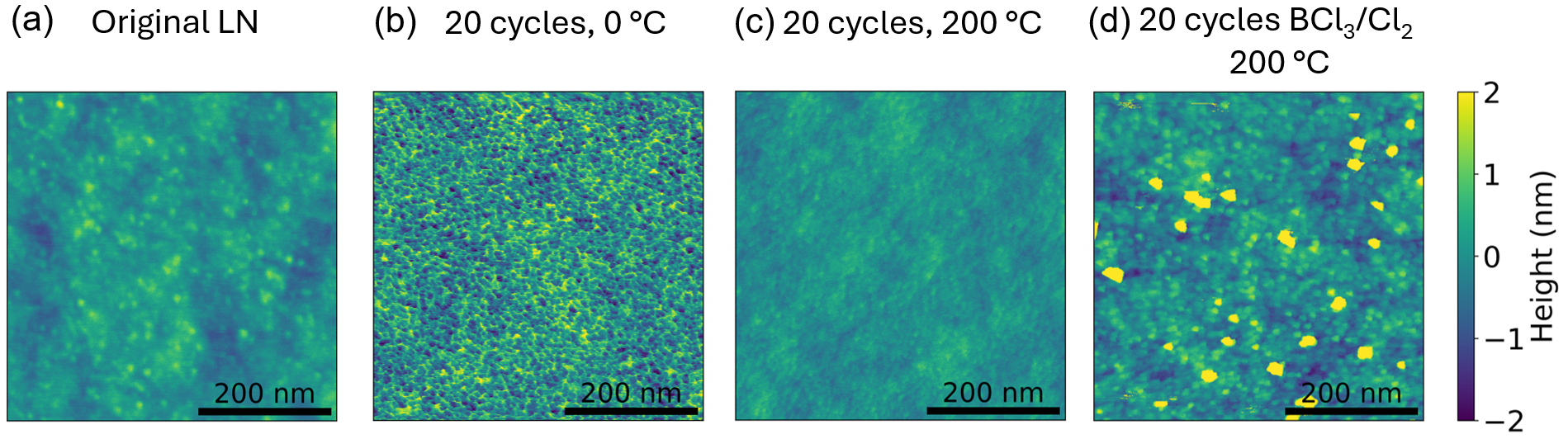}
    \phantomsubcaption\label{fig:before_ALE}
    \phantomsubcaption\label{fig:20cyc_0deg}
    \phantomsubcaption\label{fig:20cyc_200deg}
    \phantomsubcaption\label{fig:Cl_20cyc_200deg}
    
}
\caption{AFM scan showing height-maps of LN surface (a) before, (b) after 20 ALE cycles at \SI{0}{\degreeCelsius}, (c) after 20 ALE cycles at \SI{200}{\degreeCelsius}, and (d) 20 ALE cycles at \SI{200}{\degreeCelsius} using a Cl$_2$/BCl$_3$ modification step. 
}
\label{fig:afm} 
\end{figure}

\Cref{fig:before_ALE,fig:20cyc_0deg} shows AFM scans before and after 20 ALE cycles at \SI{0}{\degreeCelsius} and \SI{200}{\degreeCelsius}. The surface roughness of the original polished surface is measured as $R_q=0.30\pm0.04$ nm. After 20 cycles at \SI{0}{\degreeCelsius}, the surface roughness is found to increase to $R_q=0.73\pm0.06$ nm. The XPS and AFM suggest that redeposition of nonvolatile etch products occurs during the etch process at \SI{0}{\degreeCelsius}. In contrast, the surface roughness after 20 cycles at \SI{200}{\degreeCelsius} is measured as $R_q=0.25\pm0.03$ nm. This value is unchanged from that of the control sample to within experimental uncertainty. The smooth surface after etching suggests that previously non-volatile compounds observed in the \SI{0}{\degreeCelsius} process are better removed at \SI{200}{\degreeCelsius}.

 To assess whether HBr chemistry facilitates desorption of non-volatile products, the HBr/BCl$_3$/Ar modification step was replaced with Cl$_2$/BCl$_3$ (20 sccm Cl$_2$, 40 sccm BCl$_3$) with all other process parameters maintained constant. This Cl$_2$/BCl$_3$ process yielded an etch rate of 1.66 $\pm 0.06$ nm/cycle and a synergy of 48.2\%. While the synergy of the HBr-based process is qualitatively compatible with that of the Cl$_2$/BCl$_3$ process, surface roughening was observed for the Cl$_2$/BCl$_3$ process as shown in \Cref{fig:Cl_20cyc_200deg}. To test if the HBr-based process  at \SI{200}{\degreeCelsius} could roughen the surface  if performed for a large enough number of cycles, we applied 120 cycles (149 nm etch) of the HBr process to a sample. The resulting surface roughness was $R_q = 0.28 \pm 0.03$ nm, and no roughening was observed. We therefore conclude that the combination of HBr chemistry and higher process temperature enhances the volatility of etch products compared to Cl etch chemistries.
 


\begin{figure}
    \centering
    
{\includegraphics[width = 450pt]{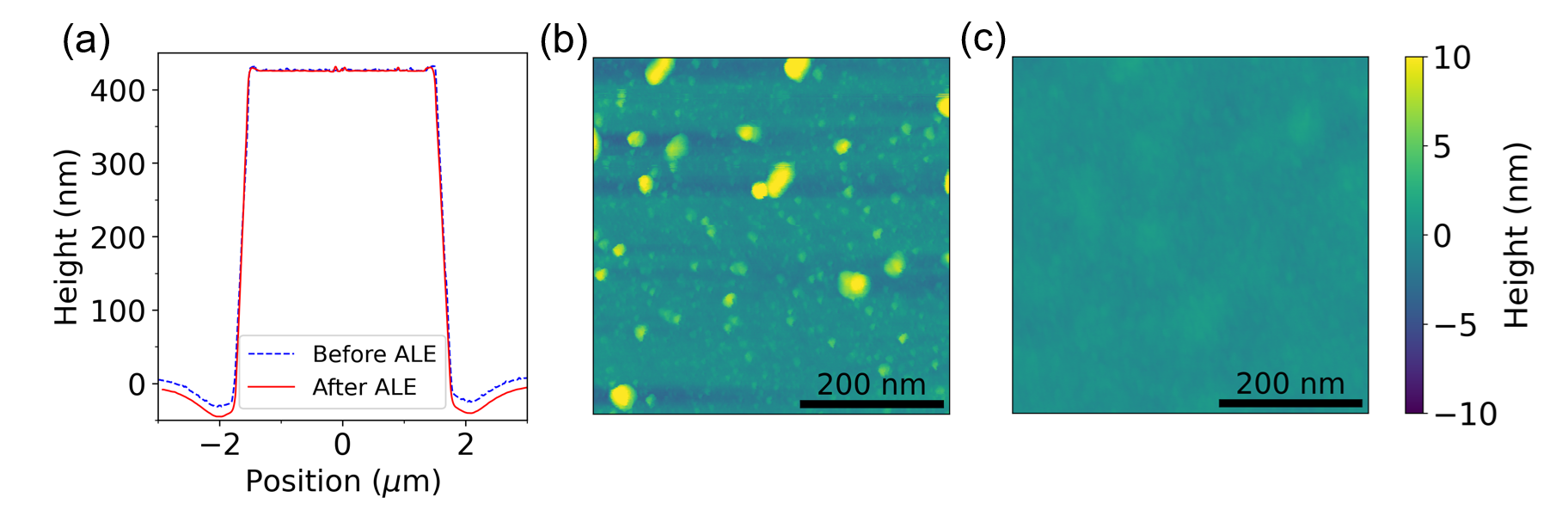}
    \phantomsubcaption\label{fig:wg-profile-directionalALE}
    \phantomsubcaption\label{fig:Armill-before}
    \phantomsubcaption\label{fig:Armill-after}
    
}
\caption{AFM scans showing (a) profile of a waveguide fabricated in TFLN using Ar ICP etching before and after directional ALE; (b) after Ar ICP etching (R$_q$ = 2.07 $\pm$ 1.16 nm, R$_a$ = 1.13 $\pm$ 0.72 nm); and (c) after 50 cycles ALE (R$_q$ = 0.34 $\pm$ 0.07 nm, R$_a$ = 0.27 $\pm$ 0.05 nm). The top width of the waveguide is maintained after ALE. The surface after HBr-based ALE is smoother than after Cl-based ALE and approaches the as-received surface roughness of unetched TFLN samples. 
}
\label{fig:wg-afm} 
\end{figure}

To assess the \SI{200}{\degreeCelsius} process for directionality and surface smoothing, 50 cycles of ALE were performed on an x-cut TFLN chip, consisting of a 990 nm crystalline x-cut LN thin film on a sapphire substrate (NanoLN), with waveguide structures fabricated by physical Ar ICP etching. The results are shown in \Cref{fig:wg-afm}. In \Cref{fig:wg-profile-directionalALE}, the waveguide profile taken by AFM before and after 50 cycles of ALE is shown. The top width of the waveguide is maintained, which indicates minimal lateral etching. We note that the microtrench depths across a waveguide profile in TFLN often differ by a few nanometers, which we speculate to arise from differential charging of the +Z and -Z faces of the waveguide sidewalls. \Cref{fig:Armill-before} and \Cref{fig:Armill-after} show AFM surface scans of the chip before and after 50 cycles of ALE. The RMS roughness of the surface was smoothed from $R_q=2.07\pm 1.16$ nm to $R_q=0.34\pm 0.07$ nm, an 84\% reduction in surface roughness. Due to the directional nature of the etch, the sidewall surface roughness of the waveguides did not decrease.

\subsection{ALE processing of TFLN device structures}
\begin{figure}
    \centering
    
{\includegraphics[width = 450pt]{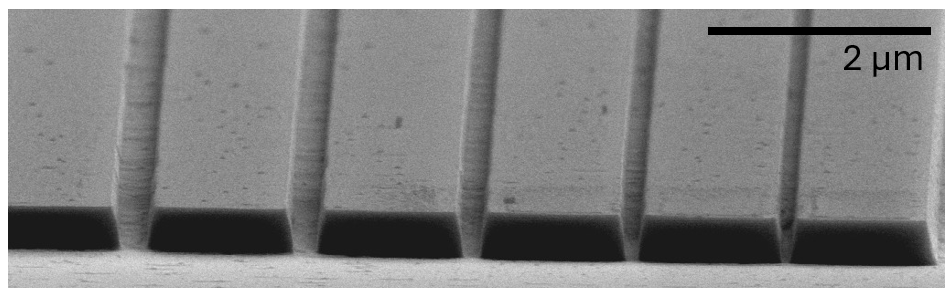}
    
}
\caption{SEM image of grating structure. The largest gap (left) in the image is 350 nm wide with decreasing gap size in increments of 50 nm. The smallest gap (right) is 150 nm wide.
}
\label{fig:SEM} 
\end{figure}
A key potential benefit of ALE for pattern transfer is its aspect ratio independent etching behavior due to the self-limiting half reactions, compared to ion milling in which the etch rate is proportional to the ion flux. Aspect ratio independent etching facilitates uniform etching in high aspect-ratio gaps, which is challenging for standard physical Ar$^+$ ICP or ion milling. To assess the aspect ratio dependence of our process etch at \SI{200}{\degreeCelsius}, a patterned 600 nm x-cut LN thin film on a SiO\textsubscript{2}/Si substrate (NanoLN) was etched with 200 cycles of directional ALE at \SI{200}{\degreeCelsius}. The pattern was a grating consisting of 19 waveguides of top width 1.2 $\mu$m with gap size decreasing from 1 um to 150 nm in increments of 50 nm. After the etch, the remaining resist was removed with a buffered HF dip. No further cleaning was performed. The total etch depth was measured via AFM as 223 nm, corresponding to an EPC of $1.12$ nm/cycle. The slightly lower etch rate from the previously measured $1.25$ nm/cycle is attributed to incomplete chamber seasoning after a manual chamber clean. \cite{samsung_seasoning} 

The sidewall angle after 200 cycles was 68 degrees. After AFM of the structures, 10 nm of carbon was deposited onto the sample to enable SEM imaging of the electrically-insulating sample. \Cref{fig:SEM} shows an SEM image of the waveguides with gap size from 150 nm to 350 nm. Despite the \SI{200}{\degreeCelsius} directional ALE process having a reduced synergy compared to the \SI{0}{\degreeCelsius} case, aspect ratio independent etching behavior is observed in that all gaps are clearly etched regardless of aspect ratio.  In comparison, Fig. 4A of Ref. \cite{KaufmannFincoMaederGrange2023} shows a FIB cross-section of a similar grating fabricated with an ion milling process of the same depth (200 nm), with aspect ratio dependent etching observed for gaps below 300 nm. The presence of stripes that span the trenches indicates that there is some redeposition into the narrow gaps. The present process may require a final process to remove any remaining redeposited products; however, there is no evidence of aspect ratio dependent etching down to the smallest tested gap of 150 nm.

\section{Discussion}

We have reported a directional ALE process using an HBr-based plasma as the modification step and a low power Ar plasma as the directional removal step. To our knowledge, this is the first reported directional ALE process for lithium niobate as well as the first to use Br-based chemistry. The process improves upon previously reported continuous and atomic layer etching approaches for LN by enhancing the volatility of the etch products, mitigating the surface roughening and premature etch termination that arise from the formation of involatile etch products. The present process is also directional, in contrast to our previous isotropic ALE process, and so will enable pattern transfer with single-nanometer etch depth precision over a wafer. 


The reported directional ALE process may find applications in LN integrated photonics by achieving single-nm level etch depth control, reduced redeposition, and overcoming the aspect ratio dependent etch physical Ar$^+$ etching. For example, coupler gratings, which are used to efficiently couple light into chips, require narrow ( $<300$ nm) gaps for high duty cycle gratings and could benefit from directional ALE as the primary etch step. \cite{Krasnokutska:19, chen2025fibertochipgratingcouplerslithium} A process consisting of the directional ALE process to obtain the desired geometric structures followed by an isotropic ALE process to smooth waveguide sidewalls can be used to etch TFLN to achieve aspect ratio independent etching, nm-level etch depth precision, and smooth waveguide sidewalls. The reported ALE processes have the potential to overcome  issues related to aspect ratio dependent etching and rough sidewalls contributing to optical losses. For periodically-poled LN (PPLN) devices like optical parametric oscillators, differential wet etch rates between poled regions from the typical wet clean causes corrugations in waveguides. Utilizing the ALE process could allow for shorter wet cleans, eliminating the formation of corrugations between different ferroelectric domains and thereby decreasing scattering losses.

Future topics of interest include incorporating the ALE process into the TFLN fabrication flow and developing an HBr-based RIE process for fast etching. To decrease the process time of ALE,  bias-pulsed ALE schemes may be of interest. \cite{Julian_pulsed_bias,DanShanks_QALE_SiN}  Further characterization of the surface species using angle resolved XPS, secondary ion mass spectrometry, and thermal desorption spectroscopy is of interest to understand the possible etch chemistries occuring in the process. Post-ALE processes like oxygen plasma ashing or annealing may be necessary to restore the surface stoichiometry and crystallinity after etching. The application of ALE to improve  quality factors of fabricated TFLN microring resonators across a chip is of interest. Development of an HBr-based RIE process in which exposures are performed simultaneously may also be of interest to achieve a faster etching rate, smoother sidewalls, and better directionality. Although an RIE process may lack the wafer-scale uniformity and aspect-ratio independent etch behavior of ALE, the RIE process would etch substantially faster than ALE.

\section{Conclusion}
We have reported a directional ALE process for lithium niobate utilizing sequential exposures of HBr/BCl$_3$/Ar and Ar plasmas. We observe an etch rate of $1.04 \pm 0.01$ nm/cycle with a synergy of 84.6\% at a temperature of \SI{0}{\degreeCelsius}. By increasing the process temperature to \SI{200}{\degreeCelsius}, synergy is decreased to 56.8\% but surface smoothing is observed. In comparison, a Cl-based process at the same temperature yields the same synergy but produces a rougher surface. An isotropic process is also reported by replacing the halogen plasma step in our previously reported process Ref. \cite{Chen2024} with the same HBr/BCl$_3$/Ar plasma in the directional process. Finally, a TFLN grating structure was etched completely by the directional ALE process at \SI{200}{\degreeCelsius} to demonstrate aspect ratio independent etching behavior down to the smallest tested gap of 150 nm. Both isotropic and directional processes could be integrated into the fabrication flow of TFLN devices to overcome the limitations of traditional physical dry etches.

\section{Acknowledgements}
This work was supported by Oxford Instruments and the Broadcom Innovation Fund at Caltech. This research was carried out, in part, at the Jet Propulsion Laboratory (JPL), California Institute of Technology, under contract with the National Aeronautics and Space Administration (NASA). We gratefully acknowledge the critical support and infrastructure provided for this work by The Kavli Nanoscience Institute and the Molecular Materials Research Center of the Beckman Institute at the California Institute of Technology. We thank Nicholas Chittock, Harm Knoops, Anthony Loveland, and Tobias Wenger for helpful discussions. We thank Ryoto Sekine for providing the bulk HSQ-patterned samples used for etch rate and synergy measurements and Richard E. Muller for patterning the TFLN HSQ-patterned sample in the aspect-ratio etch dependence study.

\section{Author Declarations}
\subsection{Conflict of Interest}
The authors have no conflicts of interest to disclose.

\section{Data Availability}
The data that support the findings of this study are available from the corresponding author upon reasonable request.

\section*{References}

\bibliographystyle{apsrev4-2}

\bibliography{./refs}

\end{document}